\documentclass[sn-mathphys,Numbered]{sn-jnl}

\usepackage{graphicx}%
\usepackage{multirow}%
\usepackage{amsmath,amssymb,amsfonts}%
\usepackage{amsthm}%
\usepackage{mathrsfs}%
\usepackage[title]{appendix}%
\usepackage{xcolor}%
\usepackage{textcomp}%
\usepackage{manyfoot}%
\usepackage{booktabs}%
\usepackage{algorithm}%
\usepackage{algorithmicx}%
\usepackage{algpseudocode}%
\usepackage{listings}%
\usepackage{hyperref}
\usepackage{xfrac}

\theoremstyle{thmstyleone}%

\theoremstyle{thmstyletwo}%

\theoremstyle{thmstylethree}%

\begin{document}

\title[Geodesic structure of a noncommutative black hole]{Geodesic structure of a noncommutative black hole}

\author[1]{\fnm{Zihan} \sur{Xi}}

\author[2]{\fnm{Chen } \sur{Wu}}

\author*[1]{\fnm{Wenjun} \sur{Guo}}\email{impgwj@126.com}

\affil*[1]{\orgname{University of Shanghai for Science and Technology}, \orgaddress{\city{Shanghai}, \postcode{200093}, \country{China}}}

\affil[2]{\orgdiv{Xingzhi College}, \orgname{Zhejiang Normal University}, \orgaddress{\city{Jinhua}, \postcode{321004}, \state{Zhejiang}, \country{China}}}

\abstract{This paper explores the metric of Piero Nicolini's noncommutative black hole spacetime, calculates its effective potential, and presents the corresponding potential curve. By analyzing this curve, we identify various orbit types for test particles and photons in this spacetime. Using the dynamical equations for particles and photons near the black hole, we plot the specific time-like and null geodesic structures. We analyze the impact of different values of the total mass of the source $M$ and angular momentum $L$ on time-like geodesics. Our results indicate that in the Piero black hole spacetime, increases in total mass and angular momentum reduce the perihelion precession rate of the orbit. Notably, the effect of total mass is nonlinear, while the effect of angular momentum is linear.}


\maketitle

\section{Introduction}\label{sec1}

Since Einstein proposed General Relativity in 1915, black holes, significant astrophysical predictions of the theory, have been focal points of research. The study of black holes has never ceased, with researchers investigating them from various perspectives to elucidate the properties of different types of black holes\cite{x1,x2,x3,x4,x5,x6,x7,x8,x9,x10,x11,x12,x13}. However, the study of black holes has faced a major challenge: singularities. At a black hole's singularity, spacetime curvature becomes infinitely large, which is physically unacceptable. Therefore, eliminating singularities within black holes has become a primary research goal. Numerous attempts have been made, with one of the most successful being Bardeen's proposal of regular black holes\cite{x14}. Since then, many singularity-free black hole solutions have been proposed\cite{x15,x16,x17}. In 2006, Piero Nicolini proposed Noncommutative Black Holes, which addressed the singularity issue by analyzing noncommutative-inspired solutions. These black holes establish a zero-temperature final state, a stable black hole remnant, with size and mass uniquely determined by the noncommutative parameter $\theta$ \cite{x18}. These findings suggest that it could resolve difficulties in traditional black hole evaporation models. Specifically, it reveals the late-stage behavior of black hole evaporation, indicates properties of the final state, and identifies an upper temperature limit, significant for understanding black hole physics and further investigating the evaporation process. Due to their characteristics and advantages, Noncommutative Black Holes have been focal points of black hole research since their proposal. Researchers conduct detailed studies on various aspects of these spacetimes. For example, Elisabetta Di Grezia studied Noncommutative Schwarzschild Black Holes and found that for finite values of $\theta$, the conformal singularity at spacelike infinity significantly reduces the differentiability class of scalar fields at future null infinity, implying slower asymptotic behavior compared to flat spacetime\cite{x19}. Kourosh Nozari investigated the tunneling process through the quantum horizon of noncommutative Schwarzschild black holes by considering a Gaussian distribution of particle mass in flat spacetime and found that due to spacetime noncommutativity, information may be preserved by a stable black hole remnant\cite{x20}. In Piero Nicolini's review, a comprehensive description known as black hole SCRAM was highlighted, referring to the cessation of black hole thermal radiation: the existence of a zero-temperature final state, a stable black hole remnant, with size and mass uniquely determined by the noncommutative parameter $\theta$\cite{x21}. Some researchers have examined noncommutative black hole spacetimes from a thermodynamic perspective. For instance, Ankur investigated charged BTZ black holes in noncommutative spacetime using two distinct approaches. Their study revealed non-static and non-stationary characteristics in noncommutative BTZ black hole spacetimes, analyzing their thermodynamic properties through contemporary tunneling formalism\cite{x22}. Rabin Banerjee derived the relationship between black hole temperature and surface gravity using tunneling interpretation and applied it to noncommutative Schwarzschild black hole spacetimes. This analysis revealed interesting temperature variation characteristics and confirmed the form of the noncommutative Bekenstein-Hawking area law\cite{x23}. In subsequent research, Rabin Banerjee and colleagues utilized image analysis to discover that for event horizon radii $r_h<4.8\sqrt\theta$ up to the extremal point $r_h=3.0\sqrt\theta$ , the standard semi-classical Bekenstein-Hawking area law holds for noncommutative Schwarzschild black holes, leading to modifications in the area law \cite{x24}. Abdellah Touati was the first to study Hawking radiation as a tunneling process within the framework of noncommutative gravity gauge theory. They discovered pure thermal radiation and logarithmic corrections to entropy. Additionally, they found that noncommutativity enhances the correlation between continuously emitted particles\cite{x25}. Other researchers have explored noncommutative black hole spacetimes from a dynamical perspective. For instance, Javlon Rayimbaev and colleagues investigated the dynamics of test particles in noncommutative black hole spacetimes. They found that while the radius of the Innermost Stable Circular Orbit (ISCO) remains unaffected by the noncommutative parameter, the frequencies of Keplerian orbits and harmonic oscillations are influenced by it. Further investigation showed that quasiperiodic oscillation orbits are located near the ISCO, potentially resolving measurement issues related to this orbit in astrophysical observations\cite{x26}. Chikun Ding analyzed the impact of noncommutative geometry on the strong field gravitational lensing effect in noncommutative Schwarzschild black hole spacetimes. They determined that this effect, while similar to that of charge, is less pronounced. This finding could help distinguish noncommutative black holes from Reissner-Nordström black holes and, in the future, might allow the noncommutative constant $\theta$ to be detected through astronomical observations\cite{x27}. Using the Anti-de Sitter/Conformal Field Theory correspondence, Xin-Yun Hu and colleagues calculated the response function of quantum field theory on the boundary of noncommutative black holes. Through a virtual convex lens optical system, they observed the Einstein ring of the black hole, noting that the ring's characteristics vary with observation positions. Depending on the observation position, the ring can appear as a luminosity-deformed ring or bright spots. Moreover, they found that as the noncommutative parameter increases, the ring's radius expands\cite{x28}. Since the proposal of noncommutative black holes, researchers have introduced an increasing variety of noncommutative black hole spacetimes, including regular, charged, rotating, higher-dimensional, and string theory-related black holes\cite{x29,x30,x31,x32,x33,x34}. The continued proposal of these black holes indicates that noncommutative black hole spacetimes have remained a popular and intriguing research topic since their inception.

Understanding geodesics is crucial for gaining detailed insights into the gravitational field near black holes. Analyzing the motion of both massive and massless particles aids in comprehending various gravitational effects around black holes. Additionally, research on geodesic motion equations explains numerous observational phenomena, such as perihelion precession, gravitational time delay, and light deflection. Consequently, geodesics, as a fundamental property of black holes, are increasingly important. In recent years, researchers have extensively studied the geodesic structures of various types of black holes\cite{x35,x36,x37}. For instance, Valeria Diemer investigated the geodesic motion of test particles in the spacetime of non-compact boson stars. She classified both massive and massless particles based on their energy and angular momentum, discussed possible orbit types, and compared her findings with observational data\cite{x38}. By providing explicit expressions for geodesic motion and conducting detailed analysis, comprehensive information about all possible orbit types and crucial features of black holes, such as the ISCO radius, can be obtained\cite{x39,x40}. This analysis uncovers many new insights. For instance, Tian Zhou demonstrated that the simplest extension of the non-analytic smooth black hole proposed by Culetu-Simpson-Visser, aside from the dual extension of radial geodesics, is geodetically incomplete in the study of geodesics of some spherically symmetric regular black holes. Similarly, the analytically extended Hayward black hole is also geodetically incomplete\cite{x41}. Furthermore, Parthapratim Pradhan conducted stability tests on circular null geodesic orbits in extremal Reissner-Nordström spacetime and discovered stable circular geodesics on the event horizon under extreme conditions. This type of geodesic is absent in the corresponding near-extremal spacetime, highlighting differences between the extremal limit of a typical Reissner-Nordström spacetime and precisely extremal geometry\cite{x42}. Kourosh Nozari and colleagues examined the Schwarzschild black hole with quantum corrections and quintessence to explore the structures of time-like and null geodesics. Their findings suggest that within this framework, quantum effects, rather than dark energy, could account for the Universe's accelerated expansion\cite{x43}. Geodesic structures provide intuitive insights into the dynamical characteristics near black holes, aiding in a clearer understanding of particle motion in gravitational fields. Consequently, researchers often include explicit diagrams of geodesic structures in their studies. These structures offer detailed information about various black hole spacetimes. For instance, Norman Cruz's research on the Schwarzschild Anti-de Sitter black hole\cite{x44}, Victor Enolskii's studies on Hořava-Lifshitz black holes \cite{x45}, and Sheng Zhou's investigations on Bardeen and Janis-Newman-Winicour spacetimes all derive conclusions by presenting geodesic equations, plotting geodesic structures, and conducting detailed analyses\cite{x46,x47}. In contemporary black hole research, this approach remains highly esteemed. Researchers like Vitor Cardoso and Brandon Bautista-Olvera have recently utilized this method in their papers, starting from geodesic equations, creating geodesic structure diagrams, and drawing specific conclusions\cite{x48,x49}. 

To explore the properties of geodesics in more detail, this paper first derives a general expression for the effective potential by analyzing the Lagrangian equations. By substituting the function $f\left(r\right)$ from the noncommutative black hole spacetime into this expression, we obtain the effective potential function. Using its differential form, we derive the dynamical orbit equations for test particles near the black hole, which are then used to plot the geodesic structures of test particles and photons. Through this analysis, we determine the characteristics and properties of geodesic structures in noncommutative black holes and assess the effects of the total mass $M$ and angular momentum $L$ on the structure of time-like geodesics.

\section{Noncommutative spherically symmetric black holes and geodesic equation}\label{sec2}
\subsection{Noncommutative spherically symmetric black holes}\label{subsec2.1}
Consider spherically symmetric and static metrics of the form

\begin{equation}
	ds^2=-f\left(r\right)dt^2+f\left(r\right)^{-1}dr^2+r^2\left(d\theta^2+\sin^2\theta d\varphi^2\right).\label{eq1}
\end{equation}

Where $r$, $\theta$, $\varphi$ and $t$ represent standard spacetime spherical coordinates, and the lapse function $f\left(r\right)$ is determined by different spacetimes. Piero introduced a novel noncommutative spherically symmetric spacetime based on noncommutative geometry\cite{x18}, with the line element given by
\begin{equation}
	ds^2=\left(1-\frac{4M}{r\sqrt\pi}\gamma\left(\sfrac{3}{2},\sfrac{r^2}{4\theta}\right)\right)dt^2-\left(1-\frac{4M}{r\sqrt\pi}\gamma\left(\sfrac{3}{2},\sfrac{r^2}{4\theta}\right)\right)^{-1}dr^2-r^2\left(d\theta^2+\sin^2\theta d\varphi^2\right).\label{eq2}
\end{equation}
The lapse function $f\left(r\right)$ is given by:
\begin{equation}
	f\left(r\right)=\left(1-\frac{4M}{r\sqrt\pi}\gamma\left(\sfrac{3}{2},\sfrac{r^2}{4\theta}\right)\right),\label{eq3}
\end{equation}
where $M$ represents the total mass of the source, $\theta$ is a constant with dimension of length squared, $\gamma\left(\sfrac{3}{2},\sfrac{r^2}{4\theta}\right)$ denotes the lower incomplete gamma function , whose specific form is given by:
\begin{equation}
\gamma\left(\sfrac{3}{2},\sfrac{r^2}{4\theta}\right)\equiv\int_{0}^{\sfrac{r^2}{4\theta}} dt \, t^{\sfrac{1}{2}} e^{-t}.\label{eq4}
\end{equation}

\subsection{Geodesic equation}\label{subsec2.2}
A static spherically symmetric metric is represented by Eq.\ref{eq1}. The corresponding Lagrangian can be expressed as:
\begin{equation}
	\mathcal{L}=-\frac{1}{2}\left[-f\left(r\right){\dot{t}}^2+f\left(r\right)^{-1}{\dot{r}}^2+r^2\left({\dot{\theta}}^2+\sin^2{\theta}{\dot{\varphi}}^2\right)\right].\label{eq5}	
\end{equation}
Where the dot indicates differentiation with respect to the affine parameter $\tau$. The Lagrangian equation is formulated as follows:
\begin{equation}
	\frac{d}{d\tau}\frac{\partial\mathcal{L}}{\partial{\dot{x}}^\nu}-\frac{\partial\mathcal{L}}{\partial x^\nu}=0.\label{eq6}
\end{equation}
Since the Lagrangian is independent of $t$ and $\varphi$, we obtain:
\begin{equation}
	\frac{\partial\mathcal{L}}{\partial\dot{t}}=-f\left(r\right)\dot{t}=-E\label{eq7}
\end{equation}	
\begin{equation}
	\frac{\partial\mathcal{L}}{\partial\dot{\varphi}}=r^2\sin^2{\theta}\dot{\varphi}=L\label{eq8}
\end{equation}	
where $E$ and $L$ are constants of motion. Choosing the initial conditions as $\theta=\frac{\pi}{2}$, we have $\dot{\theta}=0$ and $\ddot{\theta}=0$. For $\theta=\frac{\pi}{2}$ , Eq.\ref{eq8} simplifies to: 
\begin{equation}
	\frac{\partial\mathcal{L}}{\partial\dot{\varphi}}=r^2\dot{\varphi}=L.\label{eq9}
\end{equation}
Substituting Eq.\ref{eq7} and Eq.\ref{eq9} into Eq.\ref{eq5}, we obtain
\begin{equation}
	\frac{{\dot{r}}^2}{f\left(r\right)}=\frac{E^2}{f\left(r\right)}-2\mathcal{L}-\frac{L^2}{r^2}.\label{eq10}
\end{equation}
In this context, we use $\eta=2\mathcal{L}$ . It is noted that $\eta$ can take on two possible values: $\eta=0$ for massless particles and $\eta=1$ for massive particles. With these conditions, we derive the orbital equation for the equatorial plane chosen in this paper:
\begin{equation}
	{\dot{r}}^2=E^2-f\left(r\right)\left(\eta+\frac{L^2}{r^2}\right).\label{eq11}
\end{equation}
Defining $f\left(r\right)\left(\eta+\frac{L^2}{r^2}\right)$ as the effective potential $V_{eff}^2$ , Eq.\ref{eq11} can be rewritten as ${\dot{r}}^2=E^2-V_{eff}^2$. By making the substitution $r=\frac{1}{u}$, the orbital equation can be transformed into:
\begin{equation}
\left(\frac{du}{d\varphi}\right)^2=\frac{E^2}{L^2}-\frac{f\left(\frac{1}{u}\right)\eta}{L^2}-f\left(\frac{1}{u}\right)u^2.\label{eq12}
\end{equation}

\section{Geodesic structure}\label{sec3}
\begin{figure}[h]%
	\centering
	\includegraphics[width=0.9\textwidth]{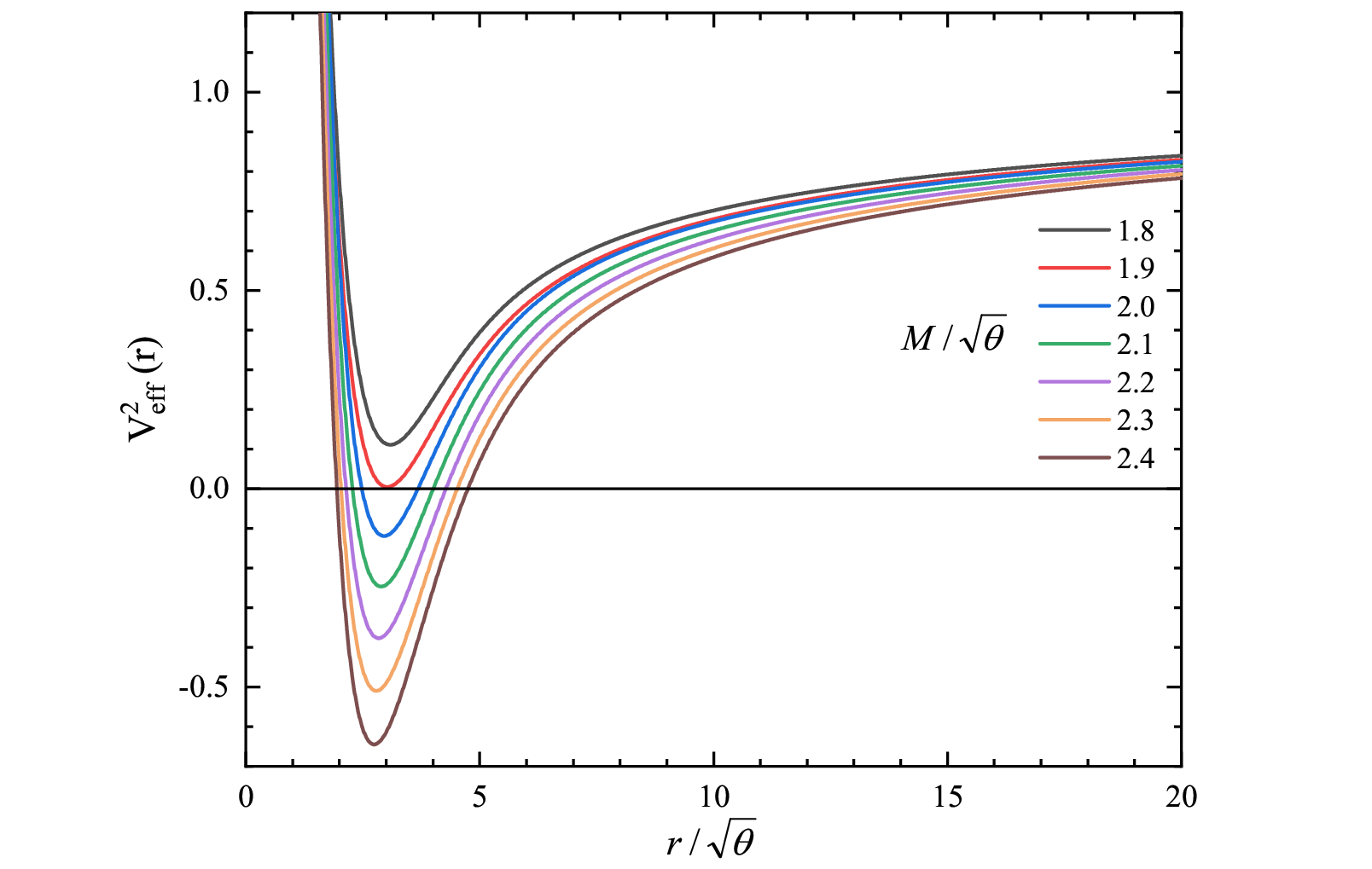}
	\caption{The effective potentials for particles with different $M$ $\left(L=3.5\sqrt\theta\right)$}\label{fig1}
\end{figure}

In the Piero spacetime, when $\eta=1$, the geodesic is time-like, and its corresponding effective potential is:
\begin{equation}
	V_{eff}^2\left(r\right)=\left(1-\frac{4M}{r\sqrt\pi}\gamma\left(\sfrac{3}{2},\sfrac{r^2}{4\theta}\right)\right)\left(1+\frac{L^2}{r^2}\right)\label{eq13}
\end{equation}
Fig.\ref{fig1} illustrates the effective potential curves for particles with varying total mass $M$ of the source in the Piero black hole spacetime, with angular momentum $L=3.5\sqrt\theta$. 

As shown in Fig.\ref{fig1} , the number of intersections between the effective potential curve and the horizontal axis corresponds to the number of horizons. When $M<1.9\sqrt\theta$, there is no horizon; when $M=1.9\sqrt\theta$, a degenerate horizon exists, indicating an extremal black hole; and when $M>1.9\sqrt\theta$, two horizons exist.

\begin{figure}[h]%
	\centering
	\includegraphics[width=0.9\textwidth]{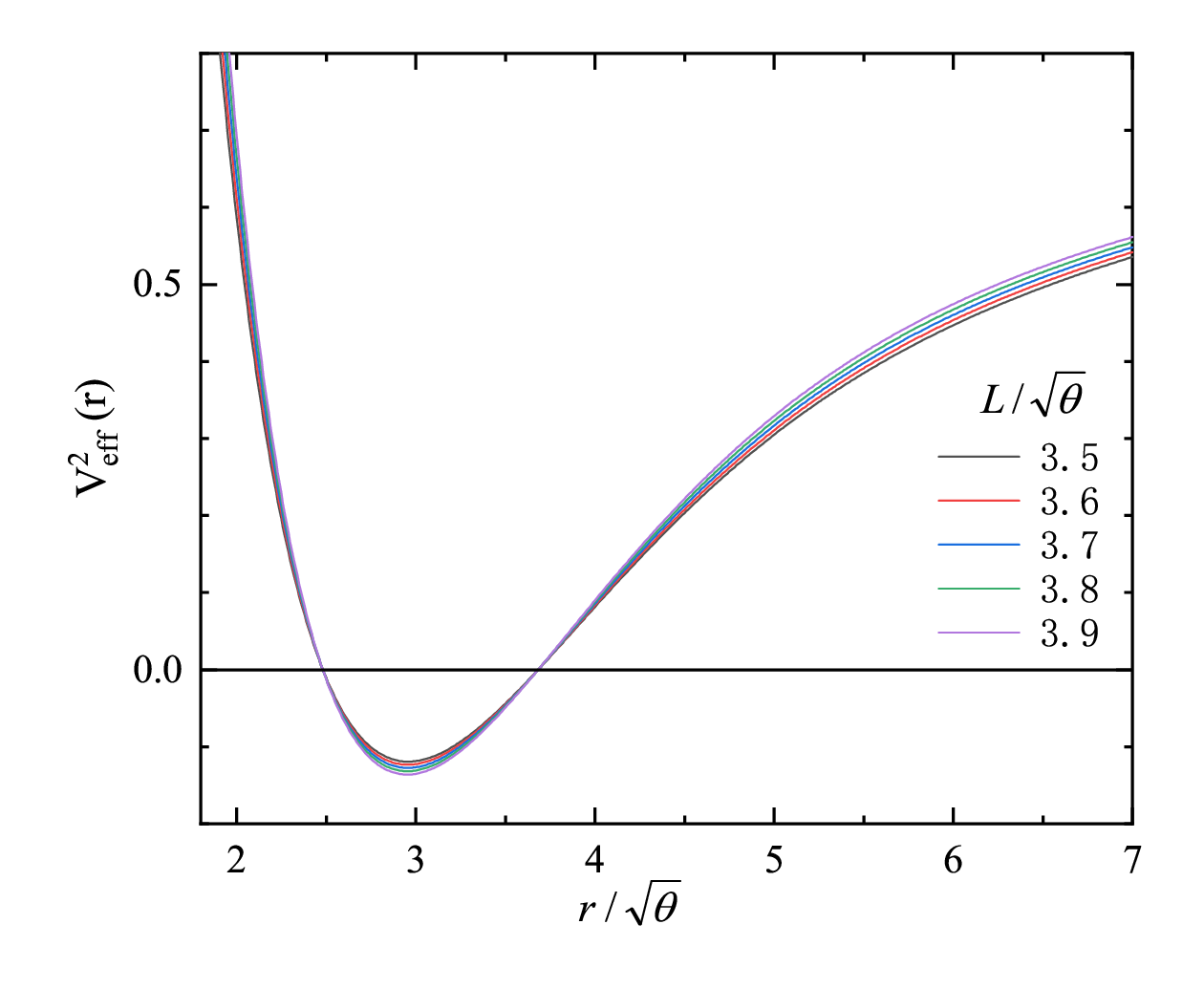}
	\caption{The effective potentials for particles with different $L \left(M=2.0\sqrt\theta\right)$}\label{fig2}
\end{figure}

Fig.\ref{fig2} illustrates the effective potential curves for particles with different angular momentum, keeping $M=2.0\sqrt\theta$ in the Piero black hole spacetime. As depicted in Fig.\ref{fig2},  it can be inferred that angular momentum does not affect the number of horizons of the black hole. In other words, regardless of the variation in angular momentum, when the total mass is fixed, the number and radius of the black hole's horizons remain unchanged. The influence of different values of total mass and angular momentum on the geodesic structure of Piero black holes will be described in detail in Section \ref{subsec3.2}.

\subsection{Time-like geodesics and null geodesics}\label{subsec3.1}

Since investigating the dynamical characteristics of particles or photons near black holes is crucial for understanding their specific properties\cite{x50}, this paper substitutes the function $f\left(r\right)$ of the Piero spacetime into Eq.\ref{eq12} and performs a second-order differentiation to derive the second-order motion equation. By numerically solving this motion equation and Eq.\ref{eq12}, we can obtain all types of time-like geodesic orbits in Piero spacetime.

\begin{figure}[h]%
	\centering
	\includegraphics[width=0.9\textwidth]{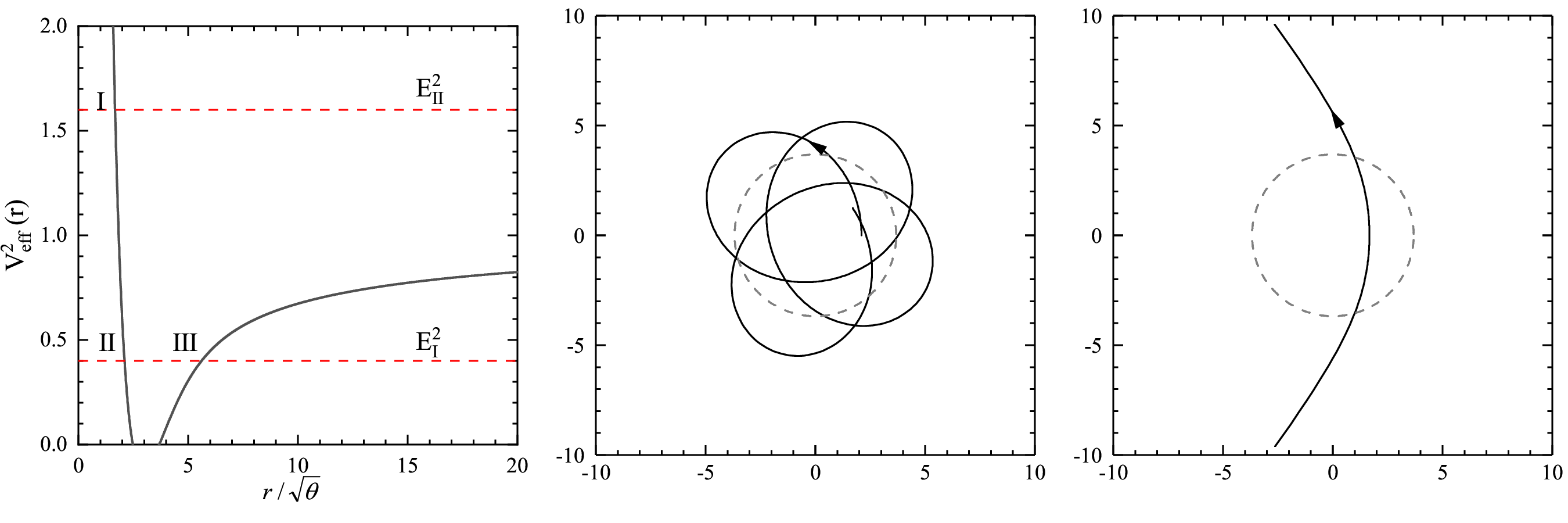}
	\caption{Time-like bound geodesics (left) and time-like escape geodesics (right) of the Piero black hole spacetime. $\left(M=2.0\sqrt\theta,L=3.5\sqrt\theta\right)$}\label{fig3}
\end{figure}

When $\eta=1$, as shown in Fig.\ref{fig3}, the intercepts of the effective potential curve with the red dashed line $E_{\uppercase\expandafter{\romannumeral1}}^2$ correspond to the radii of periapsis and apoapsis, while the gray dashed line represents the outer event horizon of the black hole. For particles with different energies, there are two types of motion: time-like bound orbits and time-like escape orbits. When $E^2<1$, the particle's orbit is bound; when $E^2>1$, the particle's orbit is an escape orbit. When $E^2=E_{\uppercase\expandafter{\romannumeral1}}^2=0.4$, the particle's orbit is bound, as depicted in the left graph of Fig.\ref{fig3}. The particle continuously crosses the horizon and remains bound near the black hole, unable to escape. At this point, the precession direction of the particle's periapsis is clockwise, opposite to the direction of motion. When $E^2=E_{\uppercase\expandafter{\romannumeral2}}^2=1.6$, the particle's orbit is an escape orbit, as shown in the right graph of Fig.\ref{fig3}. The particle enters from infinity, approaches the black hole, its orbit deflects, enters the horizon, bends further, and finally exits the horizon, returning to infinity along a straight line to complete its escape.

\begin{figure}[h]%
	\centering
	\includegraphics[width=0.9\textwidth]{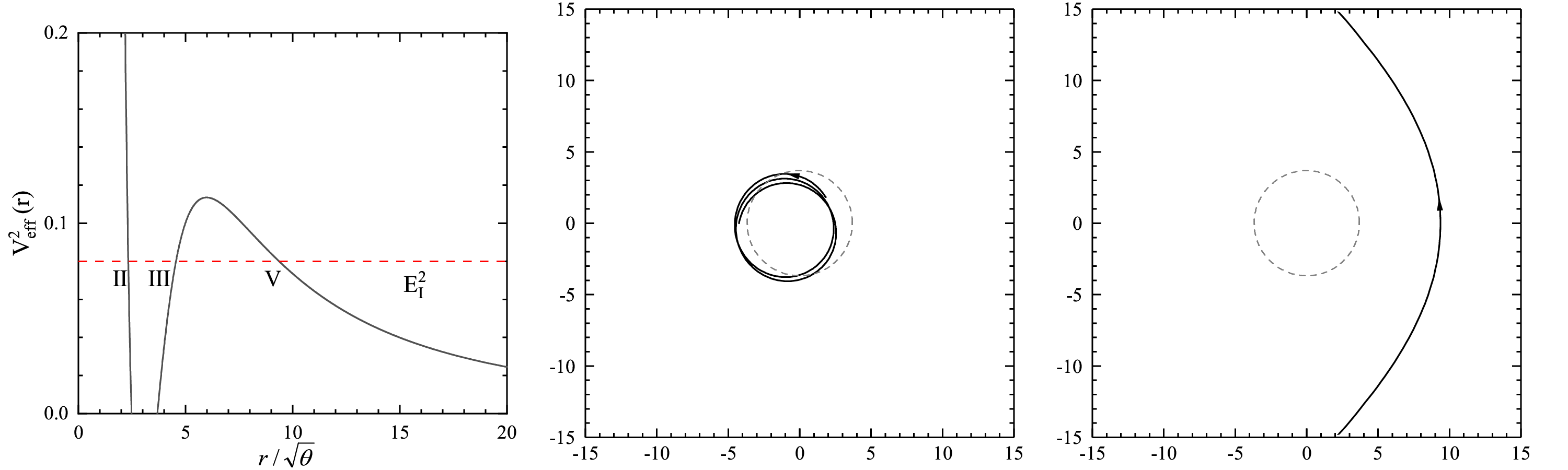}
	\caption{Null bound geodesics (left) and null escape geodesics (right) of the Piero black hole spacetime.$\left(M=2.0\sqrt\theta,L=3.5\sqrt\theta\right)$}\label{fig4}
\end{figure}

\begin{figure}[h]%
	\centering
	\includegraphics[width=0.9\textwidth]{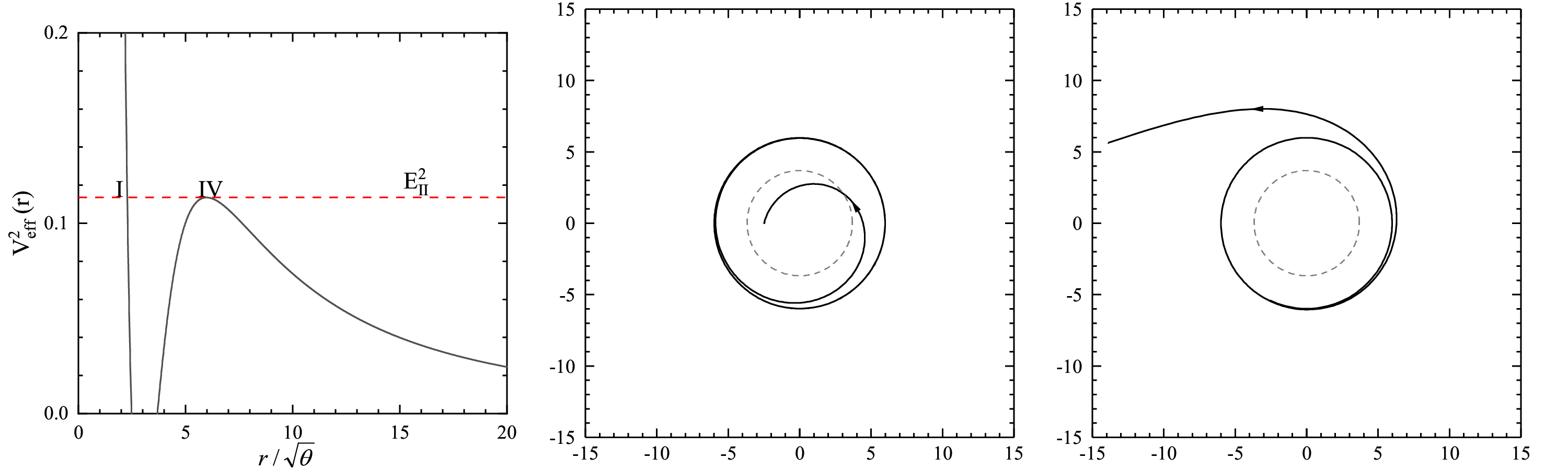}
	\caption{Unstable null circle geodesics of the Piero black hole spacetime.$\left(M=2.0\sqrt\theta,L=3.5\sqrt\theta\right)$}\label{fig5}
\end{figure}

When $\eta=0$, as shown in Fig.\ref{fig4} and Fig.\ref{fig5}, $E_{\uppercase\expandafter{\romannumeral2}}^2$ represents the critical energy value. For photons with varying energies, three distinct types of motion orbits exist. When $E^2<E_{\uppercase\expandafter{II}}^2$, photon orbits include null bound orbits and null escape orbits, as shown in the left panel of Fig.\ref{fig4}. The periapsis and apoapsis radii of the bound orbit are determined by the abscissas of points II and III. Similar to particles, photons are confined near the black hole, continuously crossing in and out of the horizon, unable to escape, and their periapsis precession speed is relatively slow. As shown in the right panel of Fig.\ref{fig4}, at this stage, photons are emitted from infinity and approach the black hole. After their trajectory deflects, the photons do not enter the horizon but travel along a curved path, eventually escaping in a straight line to infinity. When the photon's energy approaches the critical energy $E^2<E_{\uppercase\expandafter{\romannumeral2}}^2$, photons move on an unstable circular orbit. In this scenario, the photon continues circular motion, but once a slight disturbance occurs, its orbit changes. Two scenarios may arise: as shown in the left panel of  Fig.\ref{fig5}, the photon orbit may become an unstable null circular bound orbit. In this case, the photon is confined near the black hole, with the periapsis and apoapsis radii determined by the abscissas of points I and IV. The photon enters the horizon and continuously moves around nearby. In the other scenario, as shown in the right panel of Fig.\ref{fig5}, the photon orbit may become an unstable null circular escape orbit. The photon's periapsis radius is the same as the circular orbit radius. Subsequently, the photon escapes along a curved path and ultimately moves in a straight line to infinity, no longer influenced by the black hole.

\subsection{Influence of different parameters on the structure of time-like geodesics}\label{subsec3.2}

The geodesic structure of test particles varies with different parameter choices\cite{x51}. To examine how varying parameters affect the geodesic structure, this study employs the effective potential depicted in Fig.\ref{fig1} to plot time-like bound geodesics at a constant energy level. By keeping the angular momentum fixed and varying the total mass from $2.0\sqrt\theta$ to $2.4\sqrt\theta$, the resulting changes in the geodesic structure are analyzed.

\begin{figure}[h]%
	\centering
	\includegraphics[width=0.9\textwidth]{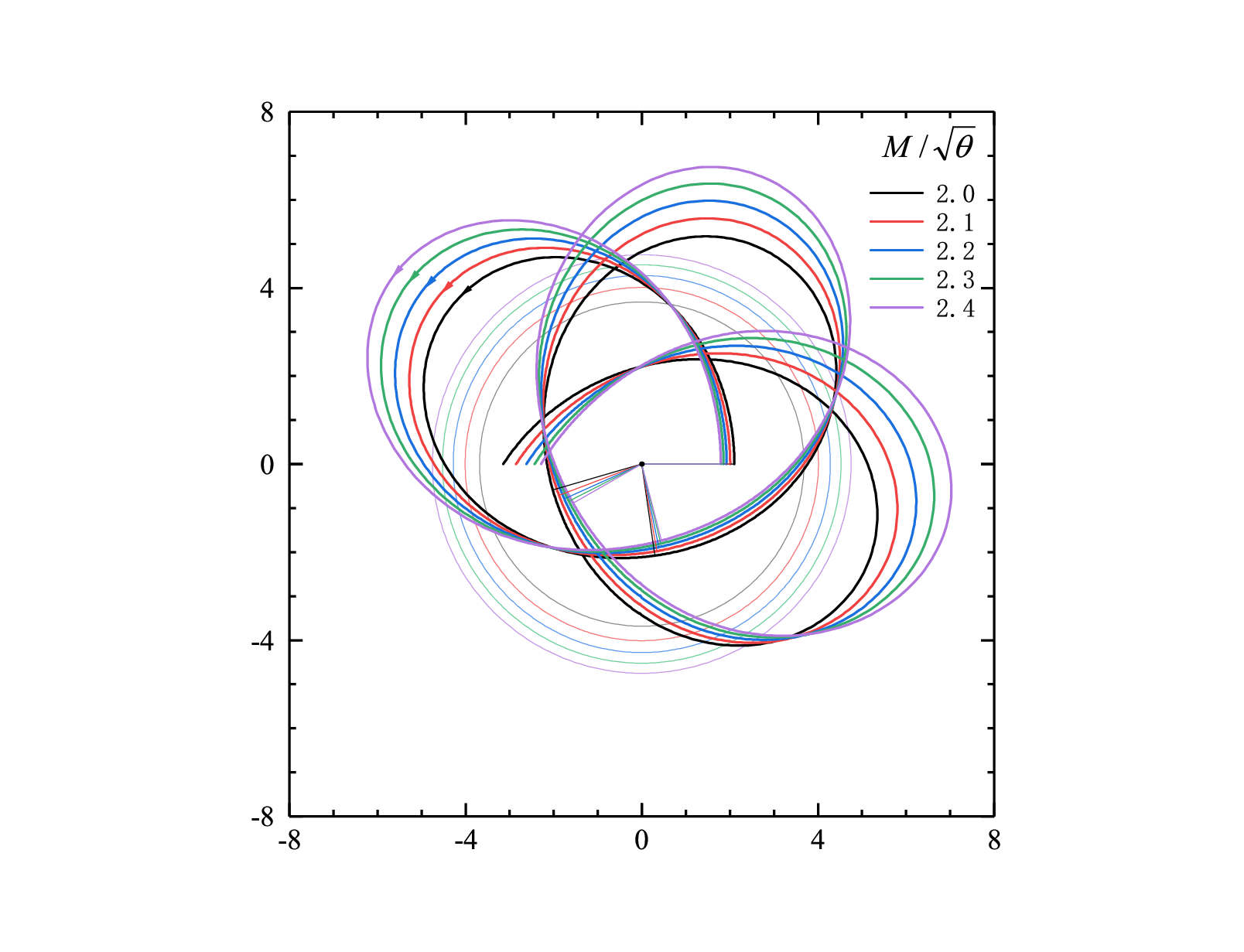}
	\caption{Time-like bound geodesics with varying $M$ in the Piero black hole spacetime. $\left(L=3.5\sqrt\theta\right)$}\label{fig6}
\end{figure}

As shown in Fig.\ref{fig6}, the different colors in the diagram represent varying values of total mass, where the darker regions depict time-like bound geodesics and the lighter regions represent the outer event horizon of the black hole for respective total mass values. It can be observed that as the total mass increases, the outer event horizon of the black hole expands accordingly. At this point, with $L=3.5\sqrt\theta$, the orbit type is a time-like bound orbit, where particles are confined near the black hole, continuously entering and exiting the event horizon but unable to escape. These orbits exhibit periodicity, with the periapsis undergoing precession. The fine straight lines, matching the colors of the bound orbits, represent periapsis annotation lines, connecting the center point to the periapsis of the orbits. Particles start from the periapsis, move in the direction of the arrows, reach the apoapsis, and eventually return to the periapsis, repeating this process continuously. The periapsis annotation lines clearly show the precession direction and speed: particle motion is counterclockwise, while periapsis precession is clockwise. It is also evident that as the total mass increases, the periapsis precession speed decreases. The influence of increasing total mass on the rate of velocity change diminishes, as indicated by the greater gap between the black and red lines compared to the green and purple lines.

\begin{figure}[h]%
	\centering
	\includegraphics[width=0.9\textwidth]{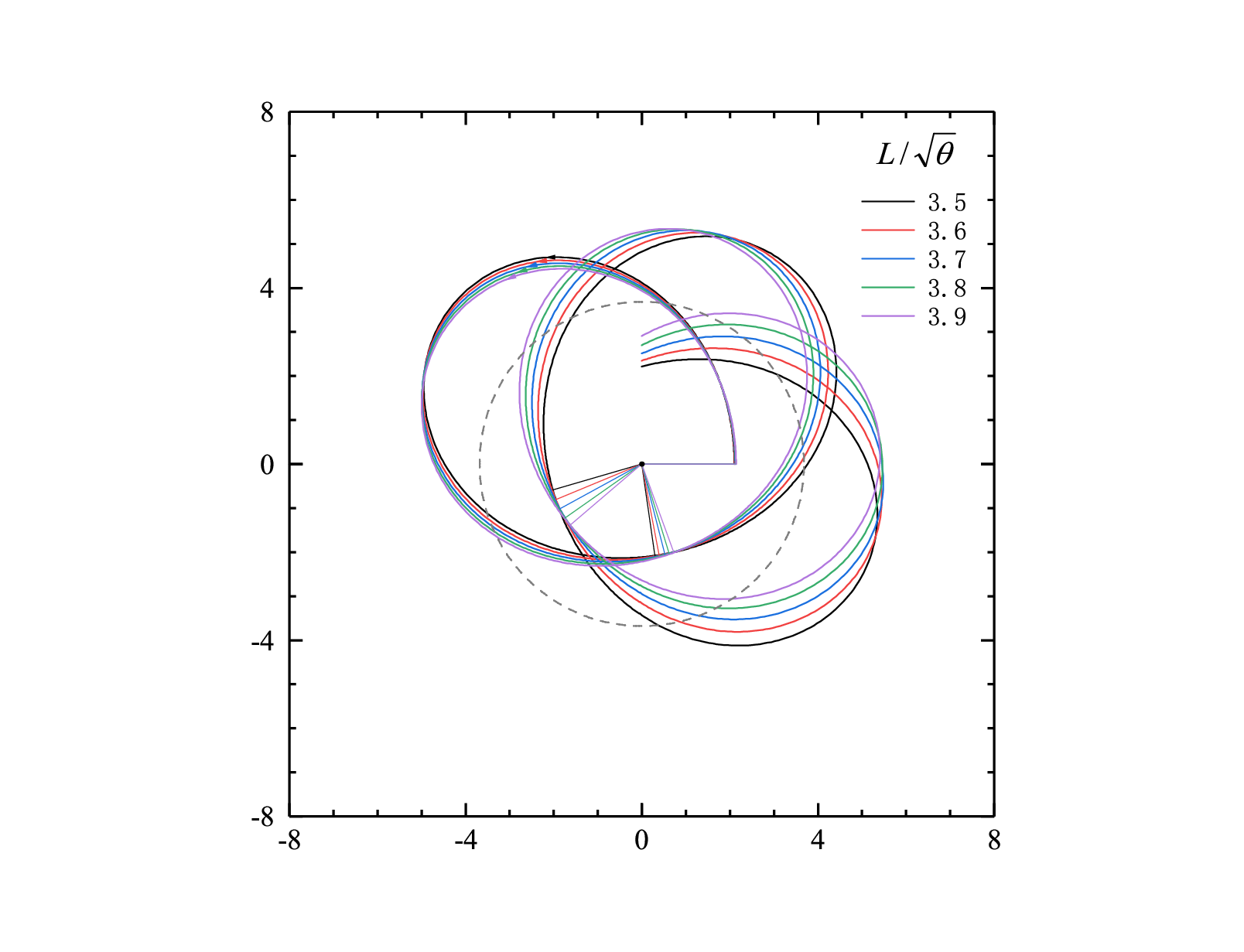}
	\caption{Time-like bound geodesics with varying $L$ in the Piero black hole spacetime. $\left(M=2.0\sqrt\theta\right)$}\label{fig7}
\end{figure}

According to the effective potential in Fig.\ref{fig2}, at the same energy level, when total mass is held constant and angular momentum increases from $3.5\sqrt\theta$ to $3.9\sqrt\theta$, the geodesic structure also changes, as shown in  Fig.\ref{fig7}.

Fig.\ref{fig7} illustrates time-like bound geodesics in the Piero spacetime with varying angular momentum, indicated by different colors, while the gray dashed line represents the outer event horizon. At this point, $M=2.0\sqrt\theta$. Particles are confined near the black hole, repeatedly crossing the event horizon during their motion but unable to escape. Starting from the periapsis, particles follow their orbit, cross the event horizon, reach the apoapsis, and return to the periapsis, completing a periodic cycle. The periapsis annotation lines indicate that the periapsis precession direction is clockwise, opposite to the direction of particle motion. As angular momentum increases, the periapsis precession speed decreases. However, the influence of angular momentum on the rate of velocity change remains linear, whereas the influence of total mass on the rate of velocity change is nonlinear.

\section{CONCLUSION}\label{sec4}

This paper investigates the Piero noncommutative black hole spacetime \cite{x18}. Starting from the general line element of a spherically symmetric black hole, we derive the corresponding Lagrangian through the principle of variation. By determining the conserved quantities and setting $\theta=\frac{\pi}{2}$ , we obtain the effective potential and the orbit equation for the equatorial plane. Through further differentiation, we derive the second-order orbit equation. 

Analyzing the effective potential of the noncommutative black hole spacetime, we identify all orbit types: bound orbits, escape orbits, and unstable circular orbits. Using the second-order orbit equation, we provide detailed images of these orbits. Our analysis shows that particles can enter and exit the event horizon in all cases, while photons can only do so in bound orbits and unstable circular bound orbits. In other scenarios, photons cannot enter or exit the event horizon. We demonstrate that in bound orbits, neither particles nor photons can escape the vicinity of the black hole. In escape orbits, particles or photons can escape to infinity. In the case of unstable circular orbits, any slight disturbance affects the orbit type of the photon and its ability to escape the black hole. 

By keeping either angular momentum or total mass constant and varying the other, we explore the impact of these parameters on the structure of time-like geodesics. Our analysis reveals that changes in total mass and angular momentum affect the precession speed of the periapsis of the orbits. As total mass increases, the precession speed of the periapsis decreases, and the rate of change in speed diminishes. When angular momentum increases, the precession speed of the periapsis also decreases, but the rate of change in speed remains relatively constant. This indicates a nonlinear relationship for total mass and a linear relationship for angular momentum regarding their impact on the rate of change in speed.

\bibliography{sn-bibliography.bib}

\end{document}